\documentclass[showpacs,preprintnumbers,amsmath,amssymb]{revtex4}

\usepackage{graphicx}
\usepackage{dcolumn}
\usepackage{bm}

\newcommand{\bq}{\begin{equation}}
\newcommand{\eq}{\end{equation}}
\newcommand{\bqa}{\begin{eqnarray}}
\newcommand{\eqa}{\end{eqnarray}}
\newcommand{\nn}{\nonumber \\}

\def\be     {\begin{equation}}
\def\ee     {\end{equation}}
\def\bea        {\begin{eqnarray}}
\def\eea        {\end{eqnarray}}
\def\bnn    {\begin{eqnarray*}}
\def\enn    {\end{eqnarray*}}

\begin{document}

\title{Kondo physics in the algebraic spin liquid}
\author{Ki-Seok Kim and Mun Dae Kim}
\affiliation{ School of Physics, Korea Institute for Advanced
Study, Seoul 130-012, Korea }
\date{\today}

\begin{abstract}
We study Kondo physics in the algebraic spin liquid, recently
proposed to describe $ZnCu_{3}(OH)_{6}Cl_{2}$ [Phys. Rev. Lett.
{\bf 98}, 117205 (2007)]. Although spin dynamics of the algebraic
spin liquid is described by massless Dirac fermions, this problem
differs from the Pseudogap Kondo model, because the bulk physics
in the algebraic spin liquid is governed by an interacting fixed
point where well-defined quasiparticle excitations are not
allowed. Considering an effective bulk model characterized by an
anomalous critical exponent, we derive an effective impurity
action in the slave-boson context. Performing the
large-$N_{\sigma}$ analysis with a spin index $N_{\sigma}$, we
find an impurity quantum phase transition from a decoupled
local-moment state to a Kondo-screened phase. We evaluate the
impurity spin susceptibility and specific heat coefficient at zero
temperature, and find that such responses follow power-law
dependencies due to the anomalous exponent of the algebraic spin
liquid. Our main finding is that the Wilson's ratio for the
magnetic impurity depends strongly on the critical exponent in the
zero temperature limit. We propose that the Wilson's ratio for the
magnetic impurity may be one possible probe to reveal criticality
of the bulk system.
\end{abstract}


\maketitle

\section{Introduction}

Recent experiments have claimed the emergence of spin liquid (SL)
phases in materials of geometrically frustrated lattices such as
$Cs_{2}CuCl_{4}$,\cite{Cs_{2}CuCl_{4}}
$\kappa-(ET)_{2}Cu_{2}(CN)_{3}$,\cite{kappa-(ET)_{2}Cu_{2}(CN)_{3}}
and $ZnCu_{3}(OH)_{6}Cl_{2}$,\cite{ZnCu_{3}(OH)_{6}Cl_{2}}, where
no symmetries associated with spin rotations (magnetic ordering)
and lattice translations (valance bond ordering) are broken at low
temperatures while charge fluctuations are frozen due to strong
electron-electron interactions (Mott insulator). An important
issue is the nature of such SL phases. Although spin
susceptibility, specific heat, and thermal transport measurements
can determine  possible spin liquids, there still remains uncertainty.

Consider $Cs_{2}CuCl_{4}$ with an anisotropic triangular
lattice.\cite{Cs_{2}CuCl_{4}} Although this material exhibits
magnetic long-range spiral ordering below $T = 0.62 K$ with an
incommensurate wave vector, the spin-fluctuation spectrum in
inelastic neutron scattering experiments has shown large
high-energy continuum beyond the spin-wave description. In
addition, this continuum spectrum survives above the Neel
temperature. More detailed analysis revealed that the continuum
follows $Im \chi(\omega) \sim \omega^{-\eta}$ with an anomalous
exponent $\eta$, suggesting the presence of deconfined critical
spinons. Such spin-fluctuation measurements  suggest several
candidates of SL scenarios, for example, decoupled one-dimensional
chains,\cite{Tsvelik} proximate gapped SLs,\cite{YBKim} algebraic spin liquid (ASL),\cite{Wen}
algebraic vortex liquid,\cite{Fisher} and so on.\cite{Senthil}

Recently, Florens {\it et al.} studied the role of magnetic
impurities in both the Z$_{2}$ SL phase and the O(4) QCP
separating the spiral magnetic order from the Z$_{2}$
SL.\cite{Florens_SL_KE} Although impurity moments coupled to
spin-$1$ bosons (spin singlet-triplet excitations) in conventional
paramagnets are only partially screened even at the bulk O(3)
QCP,\cite{Sachdev_Vojta} they have shown that the presence of
deconfined bosonic spinons can display a bosonic version of the
Kondo effect. Furthermore, they found a weak-coupling impurity
quantum phase transition (I-QPT) from a local-moment state to a
fully-screened phase. This study implies that the magnetic
impurity can be utilized as a probe for elementary excitations,
thus identifying the nature of SLs.

In this paper we investigate the Kondo effect in the ASL, recently
proposed to be realized in the Kagome antiferromagnet
$ZnCu_{3}(OH)_{6}Cl_{2}$,\cite{PALee_ASL} where no magnetic order
is observed down to very low temperature $50 mK$ compared to the
Curie-Weiss temperature $(> 200 K)$, and there is no sign of spin
gap in dynamical neutron scattering.\cite{ZnCu_{3}(OH)_{6}Cl_{2}}
However, it is not perfectly clear whether all experiments are
consistent with the ASL conjecture. The ASL picture is not
consistent with the temperature-linear specific heat below $0.5 K$
and saturation of the spin susceptibility to a finite value below
$0.3 K$,\cite{PALee_ASL} because these measurements indicate the
existence of a finite density of states at the Fermi energy. This
discrepancy may result from the presence of disorder in real
materials. To examine the role of magnetic impurities in the ASL
can be an important test in revealing the genuine nature of the SL
phase of this compound.

The ASL can be found from the fermion representation of the
Heisenberg model via the flux mean-field
ansats.\cite{Marston_flux} This reminds us of the previous study
of the Kondo effect in the flux phase by Cassanello and
Fradkin.\cite{Fradkin_KE_Flux} More generally, one may regard the
present impurity problem as the class of the Pseudogap Kondo
model,\cite{Fradkin_KE_Flux,Vojta_PG_KM,Ingersent_NRG} where the
fermion density of states vanishes as $\rho(\epsilon) \sim
|\epsilon|^{r}$ near the Fermi energy. The case of $r = 0$
corresponds to Fermi liquid while the $r = 1$ case coincides with
Dirac fermions arising from the flux phase or $d-wave$
superconductor. In contrast with the Kondo effect of the Fermi
liquid, the Pseudogap Kondo model has shown that Kondo screening
of the magnetic impurity can appear beyond some critical value of
the Kondo coupling constant. Thus, the I-QPT from a local-moment
state to a Kondo-screened phase was found in this model.
Furthermore, the exponent $r$ in the density of states was shown
to play the role of an effective dimension in the problem. The $r
= 1$ case was found to be its upper critical dimension, thus
exhibiting logarithmic corrections to scaling while the case of $r
= 0$ lies in its lower critical one.

However, there is an important difference between the Pseudogap
Kondo problem and ASL Kondo physics. The bulk physics in the
Pseudogap Kondo problem is governed by a noninteracting (Gaussian)
fixed point, thus allowing well-defined electron-like
quasiparticle excitations. On the other hand, the ASL physics is
determined by an interacting fixed point (the conformal invariant
fixed point of QED$_{3}$),\cite{ASL_fixed_point} where
well-defined spinon quasiparticle excitations corresponding to
electrons do not exist. The absence of quasiparticle excitations
prohibits us from applying the conventional picture of the
Pseudogap Kondo physics to the ASL Kondo problem. In this respect
the Kondo effect at such an interacting fixed point is an
interesting problem.

The main difficulty is how to introduce the absence of
well-defined spinon excitations in the ASL Kondo problem.
Long-range gauge interactions would result in the anomalous
critical exponent $\eta_{\psi}$ in the single spinon propagator,
destroying the quasiparticle pole in the Green's function.
Unfortunately, such critical physics can be found within the
summation of infinite diagrams of gauge interactions, and this
procedure prohibits us from analyzing the ASL Kondo problem in a
simple mean-field way such as the large-$N_{\sigma}$ approximation
with a spin index $N_{\sigma}$, well utilized in the Pseudogap
Kondo problem.\cite{Fradkin_KE_Flux} Considering the mathematical
derivation in the large-$N_{\sigma}$ context, the main problem is
how to derive an effective impurity action from the ASL-Kondo
Lagrangian through integrating out bulk degrees of freedom,
critical spinon and gauge fluctuations coupled to the magnetic
impurity. More precisely speaking, a bulk-spinon propagator
appears to govern the impurity dynamics in the effective impurity
action, thus how to write its accurate form is an important
problem since the presence of gauge interactions makes such a task
nontrivial.

In the present paper we assume the expression of the spinon
Green's function as an ansatz, introducing an anomalous critical
exponent $\eta_{\psi}$. In the text we discuss the validity of
this ansatz in great detail. This effective representation allows
us to analyze the ASL Kondo problem in the large-$N_{\sigma}$
context. Performing the slave-boson saddle-point analysis for the
effective impurity action, we find an I-QPT from a decoupled
local-moment state to a Kondo-screened phase. We evaluate the
impurity spin susceptibility and specific heat coefficient at zero
temperature, and find that such responses follow power-law
dependencies due to the ASL anomalous exponent. The main finding
of the present study is that the Wilson's ratio for the magnetic
impurity depends strongly on the ASL critical exponent in the zero
temperature limit. We propose that the Wilson's ratio for the
magnetic impurity be a probe to reveal criticality of the bulk
system.

\section{Review of the algebraic spin liquid and its Kondo problem}

For completeness of this paper, it is necessary to review how the
effective Lagrangian so called QED$_{3}$ describing the ASL is
derived from a microscopic model such as the antiferromagnetic
Heisenberg model, $H = \sum_{ij}J_{ij}\vec{S}_{i}\cdot\vec{S}_{j}$
with $J_{ij}>0$. Inserting the fermion representation of spin
$\vec{S}_{i} =
\frac{1}{2}\sum_{\sigma\sigma'}f^{\dagger}_{i\sigma}\vec{\tau}_{\sigma\sigma'}f_{i\sigma'}$
into the Heisenberg model, and performing the Hubbard-Stratonovich
transformation for an exchange channel, we find an effective
one-body Hamiltonian for fermionic spinons ($f_{i\sigma}$) coupled
to a hopping parameter ($\chi_{ij}$), $H_{eff} = -
\sum_{ij\sigma}J_{ij}f_{i\sigma}^{\dagger}\chi_{ij}f_{j\sigma} +
\sum_{ij}J_{ij}|\chi_{ij}|^{2}$. Notice that the hopping parameter
$\chi_{ij}$ is a complex number defined on links $ij$. Thus, it
can be decomposed into $\chi_{ij} = |\chi_{ij}|e^{i\theta_{ij}}$,
where $|\chi_{ij}|$ and $\theta_{ij}$ are the amplitude and phase
of the hopping parameter, respectively. Inserting this
representation of $\chi_{ij}$ into the effective Hamiltonian, we
obtain $H_{eff} = -
\sum_{ij\sigma}J_{ij}|\chi_{ij}|f_{i\sigma}^{\dagger}e^{i\theta_{ij}}f_{j\sigma}$,
where the constant contribution for the ground state energy is
omitted. Then, we can see that this effective Hamiltonian has an
internal U(1) gauge symmetry, $H'_{eff}[f'_{i\sigma},\theta'_{ij}]
= H_{eff}[f_{i\sigma},\theta_{ij}]$ under the following U(1) phase
transformation, $f'_{i\sigma} = e^{i\phi_{i}}f_{i\sigma}$ and
$\theta'_{ij} = \theta_{ij} + \phi_{i} - \phi_{j}$. This implies
that the phase $\theta_{ij}$ of the hopping parameter plays the
same role as the U(1) gauge field $a_{ij}$.

One can perform a saddle-point analysis of the effective
Hamiltonian to find its stable mean-field phases in various
lattices such as square,\cite{Marston_flux} triangular,\cite{Wen}
Kagome,\cite{PALee_ASL,Hastings} and etc. In the present paper we
consider the square lattice for simplicity, where the
antiferromagnetic long-range order can be suppressed via
next-nearest-neighbor or ring-exchange interactions causing
frustration. It is not so difficult to extend the mean-field
analysis on the square lattice into that on the Kagome lattice,
proposed to show the SL physics of
$ZnCu_{3}(OH)_{6}Cl_{2}$.\cite{PALee_ASL}

It has been shown that one possible stable mean field phase is a
$\pi$-flux state, where a spinon gains the phase of $\pi$ when it
turns around one plaquette. The amplitude of the hopping parameter
is frozen to be $|\chi_{ij}| = \sum_{\sigma}|\langle
f_{j\sigma}^{\dagger}f_{i\sigma}\rangle| \equiv \chi_{0}$ in the
low energy limit. Then, one finds the low-energy effective
Lagrangian in terms of massless Dirac fermions interacting via
compact U(1) gauge fields\cite{DonKim_QED} \bqa && Z =
\int{D\psi_{n\sigma}}{Da_{\mu}}e^{-\int{d^3x} {\cal L}} , \nn &&
{\cal L} = \sum_{\sigma=\uparrow,\downarrow}\sum_{n=1}^{2}
\bar{\psi}_{n\sigma}\gamma_{\mu}(\partial_{\mu} -
ia_{\mu})\psi_{n\sigma} + \frac{1}{2e^2}|\partial\times{a}|^2 .
\eqa Here, $\psi_{n\sigma}$ is the two-component massless Dirac
spinon, where $n = 1, 2$ represent the nodal points of
$(\pi/2,\pi/2)$ and $(\pi/2,-\pi/2)$, and ${\sigma} = \uparrow,
\downarrow$, SU(2) spin. They are expressed as $\psi_{1\sigma} =
\left(\begin{array}{c} f_{1e\sigma} \\ f_{1o\sigma} \end{array}
\right)$ and $\psi_{2\sigma} = \left( \begin{array}{c} f_{2o\sigma} \\
f_{2e\sigma} \end{array} \right)$, respectively. In the spinon
field $f_{nl\sigma}$ $n = 1, 2$ represent the nodal points, $l =
e, o$, even and odd sites, and $\sigma = \uparrow, \downarrow$,
its spin, respectively. The Dirac matrices $\gamma_{\mu}$ are
given by the Pauli matrices $\gamma_{\mu} = (\sigma_{3},
\sigma_{2}, \sigma_{1})$, satisfying the Clifford algebra
$[\gamma_{\mu},\gamma_{\nu}]_{+} = 2\delta_{\mu\nu}$. $a_{\mu}$ is
the U(1) gauge field whose kinetic energy results from
particle-hole excitations of high energy spinons. $e$ is an
effective internal charge, not a real electric charge.

It has been argued that QED$_{3}$ has an infrared stable fixed
point showing the conformal symmetry in the large-$N_{\sigma}$
limit ($\sigma = 1, ..., N_{\sigma}$).\cite{ASL_fixed_point} This
conformal invariant fixed point is identified with the ASL,
displaying algebraically decaying correlation functions with
anomalous critical exponents. To confirm the ASL as a genuine stable phase
a cautious person may ask the
stability of such an interacting fixed point against perturbations.
Four-fermion
interaction terms are irrelevant at this fixed point owing to
their high scaling dimensions. In addition, chiral symmetry
breaking due to noncompact gauge fluctuations has been shown not
to occur in the Schwinger-Dyson-equation analysis when the flavor
number of massless Dirac fermions is sufficiently large.\cite{CSB}
Furthermore, it has been argued that confinement as an instanton
effect arising from compact gauge fluctuations does not seem to
appear in the large-$N_{\sigma}$ limit because the scaling
dimension of the monopole insertion operator is proportional to
the flavor number $N_{\sigma}$, thus expected to be irrelevant in
the large-$N_{\sigma}$ ASL.\cite{ASL_fixed_point}

Criticality of the ASL is characterized by critical exponents of
correlation functions. The single particle propagator $G_{ASL}(k)
= \langle\psi_{n\sigma}(k)\bar{\psi}_{n\sigma}(k)\rangle$ can be
expressed as \bqa G_{ASL}(k) \approx
-i\frac{\gamma_{\mu}k_{\mu}}{k^{2-\eta_{\psi}}} , \eqa where
$\eta_{\psi}$ is an anomalous critical exponent. One can find such
an anomalous dimension in the large-$N_{\sigma}$
analysis.\cite{Exponent_Large_N} However, the critical exponent
obtained in this way is difficult to have a definite physical
meaning because it is not gauge invariant. In this respect the
critical exponent $\eta_{\psi}$ should be evaluated in a gauge
invariant way. The following gauge invariant Green's function can
be considered, $G_{ASL}(x) = \langle
T_{\tau}[\psi_{n\sigma}(x)e^{i\int_{0}^{x}d\zeta_{\mu}a_{\mu}(\zeta)}\bar{\psi}_{n\sigma}(0)]\rangle$.
Unfortunately, it is not easy to calculate the critical exponent
with such a gauge invariant expression. Its precise value
is far from consensus and still under current debates. The crucial
point is the sign of the exponent $\eta_{\psi}$ while its absolute
value is given by $|\eta_{\psi}| \sim N_{\sigma}^{-1}$ in the
$1/N_{\sigma}$ approximation.\cite{Exponent_Large_N}. Most
evaluations\cite{Wen_ARPES,Ye_propagator,Khveshchenko_propagator,QED_eta}
suggest its negative sign, $\eta_{\psi} < 0$. However, as argued
in Ref. \cite{Tesanovic_QED}, its negative sign seems to be
unphysical in the sense that the spinon propagator becomes more
"coherent" at long distances than the propagator of the free Dirac
theory. This result is in contrast with the usual role of
interactions, making elementary excitations less coherent. This is
indeed true in such critical field theories with local repulsive
interactions, for example, the $N$-vector model, where positive
critical exponents are well known.\cite{N_Vector_Model} If the
critical exponent is positive, long-range gauge interactions
destabilize the quasiparticle pole. The quasiparticle weight $Z(p)
\sim p^{\eta_{\psi}}$ with momentum $p$ vanishes in the long-wave
length and low-energy limits. In the present paper we do not
determine its sign. Instead, we regard the exponent $\eta_{\psi}$
as a phenomenological parameter. Thus, we consider both cases of
$\eta_\psi < 0$ and $\eta_\psi > 0$. Furthermore, we assume that
the renormalized spinon propagator [Eq. (2)] is obtained in a
gauge invariant
way,\cite{Wen_ARPES,Tesanovic_QED,Ye_propagator,Khveshchenko_propagator,QED_eta}
and the critical exponent $\eta_{\psi}$ is also gauge invariant.

Another important character of the ASL is that the conformally
invariant fixed point has an enlarged global symmetry beyond the
original lattice model, here the Heisenberg Hamiltonian. Such an
emergent symmetry corresponds to Sp(4) in the case of SU(2) gauge
interactions\cite{Wen_Symmetry} and SU(4) in the case of U(1)
ones\cite{Hermele_Symmetry}. This enlarged symmetry gives rise to
an important effect on correlation functions, that is, resulting
in the same behaviors between different correlation functions when
the operators in the correlators are related with symmetry
transformations. For example, staggered spin correlations have the
same functional dependency (power-law decay) as the valance bond
fluctuations since they are symmetry-equivalent. An interesting
point is that such correlations are most susceptible in the
ASL.\cite{Hermele_Symmetry} This implies that the ASL resides near
the antiferromagnetic and valance bond solid phases. Actually,
Tanaka and Hu have derived an effective Wess-Zumino-Witten (WZW)
Lagrangian from the ASL, describing competition between
antiferromagnetic spin correlations and valance bond
fluctuations.\cite{Tanaka}

To study the role of magnetic impurities in the ASL bulk, we
consider the Kondo coupling term, $H_{K} = \frac{J_{K}}{2}
\sum_{q}{\vec S}_{q}\cdot{\vec s}$, where ${\vec S}_{q}$ is a
spin-fluctuation operator of bulk spinons with momentum $q$ and
${\vec s}$ represents an impurity spin. The bulk-spin operator has
two contributions in the continuum, \bqa && {\vec S}(q) \approx
{\vec S}_{u}(q) + {\vec S}_{s}(q) =
\sum_{k}\sum_{n\sigma\sigma'}{\bar
\psi}_{n\sigma}(k-q)\gamma_{0}\frac{\vec
\tau_{\sigma\sigma'}}{2}\psi_{n\sigma'}(k) +
\sum_{k}\sum_{n\sigma\sigma'}{\bar \psi}_{n\sigma}(k-q)\frac{\vec
\tau_{\sigma\sigma'}}{2}\psi_{n\sigma'}(k) , \eqa where ${\vec
S}_{u}(q)$ represents the uniform component and ${\vec S}_{s}(q)$
denotes the staggered one.\cite{DonKim_QED} Then, the ASL Kondo
problem is described by the following action \bqa && {\cal S} =
\int{d\tau} \Bigl\{ \int d^2r \Bigl(
\sum_{\sigma=\uparrow,\downarrow}\sum_{n=1}^{2}
\bar{\psi}_{n\sigma}\gamma_{\mu}(\partial_{\mu} -
ia_{\mu})\psi_{n\sigma} + \frac{1}{2e^2}|\partial\times{a}|^2
\Bigr)  + \frac{J_{K}}{2} \sum_{q}\bigl({\vec S}_{u}(q) + {\vec
S}_{s}(q)\bigr)\cdot{\vec s} \Bigr\} . \eqa

The next work is to obtain an effective impurity action,
integrating out bulk degrees of freedom, spinon and gauge
excitations coupled to the magnetic impurity. One can write down
its schematic expression in the following way \bqa && {\cal
S}_{imp} \approx - \frac{J_{K}^{2}}{4} \int {d\tau d\tau'}
s^{a}(\tau)\Bigl(\sum_{q}\langle
{S}^{a}_{u}(q,\tau){S}^{b}_{u}(-q,\tau') \rangle + \sum_{q}\langle
{S}^{a}_{s}(q,\tau){S}^{b}_{s}(-q,\tau') \rangle
\Bigr)s^{b}(\tau') + \cdots , \eqa where $\langle
{S}^{a}_{s(u)}(q,\tau){S}^{b}_{s(u)}(-q,\tau') \rangle$ is the
renormalized correlation function of staggered (uniform) spin
fluctuations, and $\cdots$ are higher moment contributions. As
clearly shown in this expression, dynamics of impurity spin
fluctuations is governed by spin correlations of the bulk at the
impurity site. An important point is that only staggered spin
correlations exhibit an anomalous scaling behavior with a
nontrivial critical
exponent.\cite{AFL_spin_correlation,Wen_spin_correlation} Uniform
spin correlations have no anomalous scaling dimension since they
correspond to conserved
currents.\cite{DonKim_QED,AFL_spin_correlation} Correlations of
conserved currents do not have any anomalous scaling dimensions.
This means that the contribution of uniform spin fluctuations is
basically the same as the Kondo effect of the Pseudogap Kondo
model while that of staggered spin excitations will give rise to
new effects on the Pseudogap Kondo physics. Furthermore, staggered
spin fluctuations are most singular in the large-$N_{\sigma}$
ASL,\cite{Hermele_Symmetry} thus expected to contribute to the
Kondo effect dominantly. In this respect we take into account
staggered spin fluctuations only, which is an important
assumption in the present paper.

\section{Kondo physics in the algebraic spin liquid: large-$N_{\sigma}$ analysis}

Our objective is to construct a mean-field theory for the present
Kondo problem. Using the slave-boson representation, the impurity
spin is expressed as ${\vec s} =
\frac{1}{2}\sum_{\sigma\sigma'}\chi_{\sigma}^{\dagger}{\vec
\tau}_{\sigma\sigma'}\chi_{\sigma'}$, and such fermions satisfy
the constraint $\sum_{\sigma =
1}^{N_{\sigma}}\chi_{\sigma}^{\dagger}\chi_{\sigma} = Q_{\chi}$
with $Q_{\chi} = 2s$, where $s$ is spin. Inserting this expression
into Eq. (4) with Eq. (3), the Kondo coupling term becomes \bqa &&
H_{K} = -
\frac{J_{K}}{2N_{\sigma}}\sum_{qk}\sum_{n=1}^{N_{n}}\sum_{\sigma\sigma'=1}^{N_{\sigma}}{\bar
\psi}_{n\sigma}(k-q)\chi_{\sigma}\chi_{\sigma'}^{\dagger}\gamma_{0}\psi_{n\sigma'}(q)
-
\frac{J_{K}}{2N_{\sigma}}\sum_{qk}\sum_{n=1}^{N_{n}}\sum_{\sigma\sigma'=1}^{N_{\sigma}}{\bar
\psi}_{n\sigma}(k-q)\chi_{\sigma}\chi_{\sigma'}^{\dagger}\psi_{n\sigma'}(q)
\eqa in the large-$N_{\sigma}$ treatment, where the first and
second terms are associated with uniform and staggered
spin-fluctuation contributions, respectively. Since staggered spin
fluctuations will give main contributions to the ASL Kondo effect,
effects of uniform spin fluctuations are neglected in the
following.

Performing the Hubbard-Stratonovich transformation for the
Kondo-exchange channel, we find an effective ASL-Kondo action as
our starting point \bqa && {\cal S}_{eff} = \int{d^3x}
\Bigl[\Bigl( \sum_{\sigma = 1}^{N_{\sigma}}\sum_{n = 1}^{N_{n}}
\bar{\psi}_{n\sigma}\gamma_{\mu}(\partial_{\mu} -
ia_{\mu})\psi_{n\sigma} +
\frac{1}{2e^2}|\partial\times{a}|^2\Bigr) - \sum_{\sigma =
1}^{N_{\sigma}}\sum_{n = 1}^{N_{n}}
(b_{n}^{s\dagger}\chi_{\sigma}^{\dagger}\psi_{n\sigma}(0) + {\bar
\psi}_{n\sigma}(0) \gamma_{0}\chi_{\sigma}b_{n}^{s})  \Bigr] \nn
&& + \int{d\tau} \Bigl[ \sum_{\sigma =
1}^{N_{\sigma}}\chi_{\sigma}^{\dagger}(\partial_{\tau} -
h_{\sigma})\chi_{\sigma} + i\lambda(\sum_{\sigma =
1}^{N_{\sigma}}\chi_{\sigma}^{\dagger}\chi_{\sigma} - Q_{\chi}) +
\frac{N_{\sigma}}{2J_{K}} \sum_{n=1}^{N_{n}}
b_{n}^{s\dagger}\gamma_{0}b_{n}^{s} \Bigr] .   \eqa The first part
represents the ASL bulk. The second part arises from the
Hubbard-Stratonovich decoupling of the Kondo interaction term,
where $b_{n}^{s}$ is a two-component hybridization order parameter
associated with staggered bulk-spin fluctuations. Such a
hybridization order parameter is determined self-consistently in
the saddle-point analysis \bqa && \frac{N_{\sigma}}{2J_{K}}
\gamma_{0}b_{n}^{s} =
\langle\int\frac{d^2k}{(2\pi)^{2}}\sum_{\sigma=1}^{N_{\sigma}}
\chi_{\sigma}^{\dagger}\psi_{n\sigma}(k)\rangle . \eqa The third
part describes impurity-spinon dynamics, where $h_{\sigma} =
\sigma H$ is an external magnetic field and $\lambda$ is a
Lagrange multiplier field to impose the pseudo-fermion constraint.

When the bulk system is in the non-interacting fixed point
corresponding to the absence of gauge interactions, the effective
Kondo model becomes the multi-channel Pseudogap Kondo model, where
the channels come from Dirac nodes $n = 1, ..., N_{n}$. This model
was argued to show an I-QPT from a decoupled local-moment state to
an over-screened phase in the large-$N_{\sigma}$ approximation
although this analysis does not capture the over-screened Kondo
physics quite well.\cite{Fradkin_KE_Flux} On the other hand, the
present bulk system lies at the interacting fixed point
characterized by the anomalous critical exponent $\eta_{\psi}$,
where quasiparticle excitations do not exist. In this case it is
not clear whether the conventional Kondo screening picture is
applicable.

Integrating out bulk-spinon and gauge excitations, we obtain an
effective impurity action in energy-momentum space \bqa && {\cal
S}_{eff}^{imp} = \int\frac{dk_{0}}{2\pi} \Bigl[ \sum_{\sigma =
1}^{N_{\sigma}} \chi_{\sigma}^{\dagger}(ik_{0} - h_{\sigma} +
\epsilon_{\chi})\chi_{\sigma} - \sum_{\sigma =
1}^{N_{\sigma}}\sum_{n = 1}^{N_{n}}
b_{n}^{s\dagger}\chi_{\sigma}^{\dagger}
%\Bigl(\int\frac{d^2k}{(2\pi)^{2}}\frac{i\gamma_{\mu}k_{\mu}}{|k|^{2-\eta_{\psi}}} \Bigr)
\Bigl(\int\frac{d^2k}{(2\pi)^{2}} \langle
\psi_{n\sigma}(k)\bar{\psi}_{n\sigma}(k)\rangle \Bigr)
\gamma_{0}\chi_{\sigma}b_{n}^{s} \nn && +
\frac{N_{\sigma}}{2J_{K}} \sum_{n=1}^{N_{n}}
b_{n}^{s\dagger}\gamma_{0}b_{n}^{s} - \epsilon_{\chi}Q_{\chi}
\Bigr] ,   \eqa where $i\lambda$ is replaced with
$\epsilon_{\chi}$ to clarify its physical meaning. The main
question in this impurity action is how to evaluate the spinon
Green's function. As discussed intensively in the previous
section, the single particle propagator has an anomalous scaling
exponent, given by $\langle
\psi_{n\sigma}(k)\bar{\psi}_{n\sigma}(k)\rangle = -
i\gamma_{\mu}k_{\mu}/|k|^{2-\eta_{\psi}}$. This expression seems
to be consistent with Eq. (5) if "sub-leading" uniform
spin-correlation contributions are not taken into account. This is
because the critical exponent of the staggered spin-spin
correlation function is found to be twice the exponent of the
single particle propagator, i.e., $2\eta_{\psi}$ in the case of
$\eta_{\psi} < 0$.\cite{Wen_spin_correlation,AFL_spin_correlation}
Such correspondence occurs when both critical exponents are
calculated in a gauge invariant way. This correspondence was also
pointed out in Ref. \onlinecite{QED_eta}.

Such spinon excitations with an anomalous scaling exponent result
in anomalous energy-dependent (nonlocal in time) interactions for
impurity fermions, as reflected in the kernel of
$\int\frac{d^2k}{(2\pi)^{2}}\frac{i\gamma_{\mu}k_{\mu}}{|k|^{2-\eta_{\psi}}}
\equiv i \gamma_{\mu} F_{\mu}(k_{0})$. The vector function
$F_{\mu}(k_{0})$ is obtained to be \bqa && F_{\mu}(k_{0}) =
\int\frac{d^2k}{(2\pi)^{2}}\frac{k_{\mu}}{(k_{0}^{2} +
k^{2})^{1-\eta_{\psi}/2}} = k_{0}\frac{(k_{0}^{2} +
\Lambda^{2})^{\eta_{\psi}/2} -
|k_{0}|^{\eta_{\psi}}}{{2\pi}\eta_{\psi}} \delta_{\mu 0} , \eqa
thus $i\gamma_{\mu} F_{\mu}(k_{0}) = i \gamma_{0} k_{0} F(k_{0})$
with $F(k_{0}) = [(k_{0}^{2} + \Lambda^{2})^{\eta_{\psi}/2} -
|k_{0}|^{\eta_{\psi}}]/2\pi\eta_{\psi}$, where $\Lambda$ is a
momentum cutoff.

Inserting Eq. (10) into Eq. (9) and integrating over impurity
fermions in Eq. (9), we find the following expression for the
impurity free energy \bqa && F_{imp} = -
\int_{-\infty}^{\infty}\frac{dk_{0}}{2\pi}\sum_{\sigma =
1}^{N_{\sigma}} \ln\Bigl[(ik_{0} - h_{\sigma} + \epsilon_{\chi}) +
ik_{0}F(k_{0})\sum_{n=1}^{N_{n}} b_{n}^{s\dagger}b_{n}^{s} \Bigr]
+ \frac{N_{\sigma}}{2J_{K}} \sum_{n=1}^{N_{n}}
b_{n}^{s\dagger}\gamma_{0}b_{n}^{s} - \epsilon_{\chi}Q_{\chi} .
\eqa Expressing the hybridization order parameter as a
two-component spinor $b_{n}^{s\dagger} = (b_{n+}^{s\dagger}
b_{n-}^{s\dagger})$, one can find $b_{n-}^{s} = 0$ in the
saddle-point analysis. Representing the above impurity free energy
with $b_{n+}^{s} \equiv 2b$, we obtain \bqa && F_{imp} = -
\frac{N_{\sigma}}{2}\int_{-\infty}^{\infty}\frac{dk_{0}}{2\pi}
\Bigl[ \ln\Bigl( ik_{0} - H + \epsilon_{\chi} +
ik_{0}\frac{2N_{n}}{\pi\eta_{\psi}}|b|^{2}\bigl\{(k_{0}^{2} +
\Lambda^{2})^{\eta_{\psi}/2} - |k_{0}|^{\eta_{\psi}}\bigr\} \Bigr)
\nn && + \ln\Bigl( ik_{0} + H + \epsilon_{\chi} +
ik_{0}\frac{2N_{n}}{\pi\eta_{\psi}}|b|^{2}\bigl\{(k_{0}^{2} +
\Lambda^{2})^{\eta_{\psi}/2} - |k_{0}|^{\eta_{\psi}}\bigr\} \Bigr)
\Bigr] + \frac{2N_{\sigma}N_{n}}{J_{K}} |b|^{2} -
\epsilon_{\chi}Q_{\chi} . \eqa Minimizing the impurity free energy
with respect to $b$ and $\epsilon_{\chi}$, we find the
saddle-point equations giving the self-consistency \bqa &&
b\Bigl(\frac{1}{J_{K}} - \int_{0}^{\infty}\frac{dk_{0}}{2\pi}
\frac{k_{0}^{2}\frac{2}{\pi\eta_{\psi}} \bigl\{(k_{0}^{2} +
\Lambda^{2})^{\eta_{\psi}/2} - k_{0}^{\eta_{\psi}}\bigr\}\bigl[1 +
\frac{2N_{n}}{\pi\eta_{\psi}}|b|^{2}\bigl\{(k_{0}^{2} +
\Lambda^{2})^{\eta_{\psi}/2} -
k_{0}^{\eta_{\psi}}\bigr\}\bigr]}{k_{0}^{2}\bigl[1 +
\frac{2N_{n}}{\pi\eta_{\psi}}|b|^{2}\bigl\{(k_{0}^{2} +
\Lambda^{2})^{\eta_{\psi}/2} -
k_{0}^{\eta_{\psi}}\bigr\}\bigr]^{2} + \epsilon_{\chi}^{2} }
\Bigr) = 0 , \nn && \frac{Q_{\chi}}{N_{\sigma}} = -
\int_{0}^{\infty}\frac{dk_{0}}{2\pi} \frac{
2\epsilon_{\chi}}{k_{0}^{2}\bigl[1 +
\frac{2N_{n}}{\pi\eta_{\psi}}|b|^{2}\bigl\{(k_{0}^{2} +
\Lambda^{2})^{\eta_{\psi}/2} -
k_{0}^{\eta_{\psi}}\bigr\}\bigr]^{2} + \epsilon_{\chi}^{2} } .
\eqa

Since the impurity free energy is momentum-cutoff-dependent, it is
necessary to make it cutoff-independent, taking appropriate
scaling transformations for all variables. Considering the scaling
dimension of $\psi_{n\sigma}$ given by $\mbox{dim}[\psi_{n\sigma}]
= 1 + \eta_{\psi}/2$, one can find $\mbox{dim}[b] = -
\eta_{\psi}/2$ and $\mbox{dim}[J_{K}] = -1 - \eta_{\psi}$, where
$\mbox{dim}[\hat{\cal O}]$ represents the scaling dimension of an
operator $\hat{\cal O}$. Then, the scale-free impurity free energy
is obtained to be \bqa && f_{imp} \equiv \frac{F_{imp}}{\Lambda} =
- \frac{N_{\sigma}}{4\pi} \int_{-\infty}^{\infty} dx \Bigl[
\ln\Bigl( ix - h + \epsilon_{r} +
ix\frac{2N_{n}}{\pi\eta_{\psi}}|b_r|^{2}\bigl\{(x^{2} +
1)^{\eta_{\psi}/2} - |x|^{\eta_{\psi}}\bigr\} \Bigr) \nn && +
\ln\Bigl( ix + h + \epsilon_{r} +
ix\frac{2N_{n}}{\pi\eta_{\psi}}|b_r|^{2}\bigl\{(x^{2} +
1)^{\eta_{\psi}/2} - |x|^{\eta_{\psi}}\bigr\} \Bigr) \Bigr] +
\frac{2N_{\sigma}N_{n}}{J_{r}} |b_r|^{2} - \epsilon_{r}Q_{\chi} ,
\eqa where such rescaled variables are given by \bqa && b_{r} =
\frac{b}{\Lambda^{-\eta_{\psi}/2}} , ~~~~~ J_{r} =
\frac{J_{K}}{\Lambda^{-(1+\eta_{\psi})}} , ~~~~~ \epsilon_{r} =
\frac{\epsilon_{\chi}}{\Lambda} , ~~~~~ x = \frac{k_{0}}{\Lambda}
, ~~~~~ h = \frac{H}{\Lambda} . \nonumber \eqa Notice that these
scaled variables are dimensionless. Accordingly, the
self-consistent saddle-point equations read \bqa &&
\frac{1}{J_{r}} = \frac{1}{\pi^{2}}\int_{0}^{\infty}dx
\frac{\frac{x^{2}}{\eta_{\psi}} \bigl\{(x^{2} + 1)^{\eta_{\psi}/2}
- x^{\eta_{\psi}}\bigr\}\bigl[1 +
\frac{2N_{n}}{\pi\eta_{\psi}}|b_{r}|^{2}\bigl\{(x^{2} +
1)^{\eta_{\psi}/2} - x^{\eta_{\psi}}\bigr\}\bigr]}{x^{2}\bigl[1 +
\frac{2N_{n}}{\pi\eta_{\psi}}|b_{r}|^{2}\bigl\{(x^{2} +
1)^{\eta_{\psi}/2} - x^{\eta_{\psi}}\bigr\}\bigr]^{2} +
\epsilon_{r}^{2} } , \nn && \frac{Q_{\chi}}{N_{\sigma}} = -
\frac{1}{\pi}\int_{0}^{\infty}dx \frac{ \epsilon_{r}}{x^{2}\bigl[1
+ \frac{2N_{n}}{\pi\eta_{\psi}}|b_{r}|^{2}\bigl\{(x^{2} +
1)^{\eta_{\psi}/2} - x^{\eta_{\psi}}\bigr\}\bigr]^{2} +
\epsilon_{r}^{2} } . \eqa

The QCP of the I-QPT can be found with $b_{r} \rightarrow 0$ and
$\epsilon_{r} \rightarrow 0$ in the particle-hole symmetric case,
$Q_{\chi}/N_{\sigma} = 1/2$. Then, the critical renormalized Kondo
coupling constant is obtained from Eq. (15), \bqa &&
\frac{1}{J_{rc}} = \frac{1}{\pi^{2}\eta_{\psi}}\int_{0}^{\infty}dx
\bigl\{(x^{2} + 1)^{\eta_{\psi}/2} - x^{\eta_{\psi}}\bigr\} =
\frac{1}{2\pi^{3/2}\eta_{\psi}}
\frac{\Gamma\bigl(-\frac{1+\eta_{\psi}}{2}\bigr)}{\Gamma\bigl(-
\frac{\eta_{\psi}}{2}\bigr)} , \eqa as far as the ASL exponent
lies in $-1 < \eta_{\psi} < 1$. In the large $N_{\sigma}$ limit
the ASL exponent may also satisfy this condition as discussed in
Section II. In addition, this critical value is continuously
defined in the limit of $\eta_{\psi} \rightarrow \pm 0$, where the
impurity critical point\cite{PGKM} is given by \bqa &&
\frac{1}{J_{rc}} = \frac{1}{2\pi^{2}}\int_{0}^{\infty}dx
\ln\bigl(1+\frac{1}{x^{2}}\bigr) = \frac{1}{2\pi} , \eqa
consistent with the previous study.\cite{Fradkin_KE_Flux}

%%%%%%%%%%%%%%%%%%%%%%%%%%%%%%%%%%%%%%%%%%%%%%%%%%%%%%%%%%%%%%%%%%%%%%%%
%Fig.1
\begin{figure}[t]
\vspace{6cm} \includegraphics{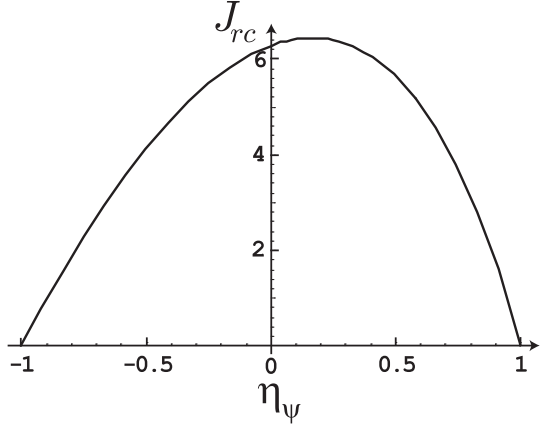} \vspace*{0cm} \caption{
Renormalized critical Kondo coupling constant as a function of the
ASL exponent. } \label{fig1}
\end{figure}
%%%%%%%%%%%%%%%%%%%%%%%%%%%%%%%%%%%%%%%%%%%%%%%%%%%%%%%%%%%%%%%%%%%%%%%%

It is interesting to notice that the I-QPT occurs as long as the
ASL exponent $|\eta_{\psi}| < 1$. Remember that in the regime of
$0 < \eta_{\psi} < 1$ critical spinon excitations are less
coherent than those in the Pseudogap Kondo model ($\eta_{\psi} =
0$) while in the regime of $-1 < \eta_{\psi} < 0$ such spinon
excitations become more coherent than quasiparticle excitations in
the Fermi liquid with pseudogap. To screen the magnetic impurity,
stronger Kondo couplings would be required when quasiparticle
excitations are less coherent. Actually, we find such an
asymmetric behavior for the ASL exponent in Fig. \ref{fig1},
obtained from Eq. (16).

It might seem mysterious that the critical Kondo coupling vanishes as
$\eta_{\psi} \rightarrow \pm 1$. As the ASL exponent approaches
$1$, critical spinon excitations are not only less coherent but
also localized. Considering the spinon propagator Eq. (2),
$\eta_{\psi} = 1$ makes it energy-momentum-independent. Such
localized spinons are expected to form a Kondo singlet with an
impurity spin immediately. When the ASL exponent goes to $-1$, it
is important that the bare scaling dimension of the Kondo coupling
($\mbox{dim}[J_{K}] = - 1 - \eta_{\psi}$) vanishes, implying that
Kondo interactions are marginal perturbations similar to the
conventional Kondo effect in the Fermi liquid. In this respect the
critical Kondo coupling would go to zero as $\eta_{\psi}
\rightarrow -1$.

\begin{figure}[t]
\vspace{6cm} \includegraphics{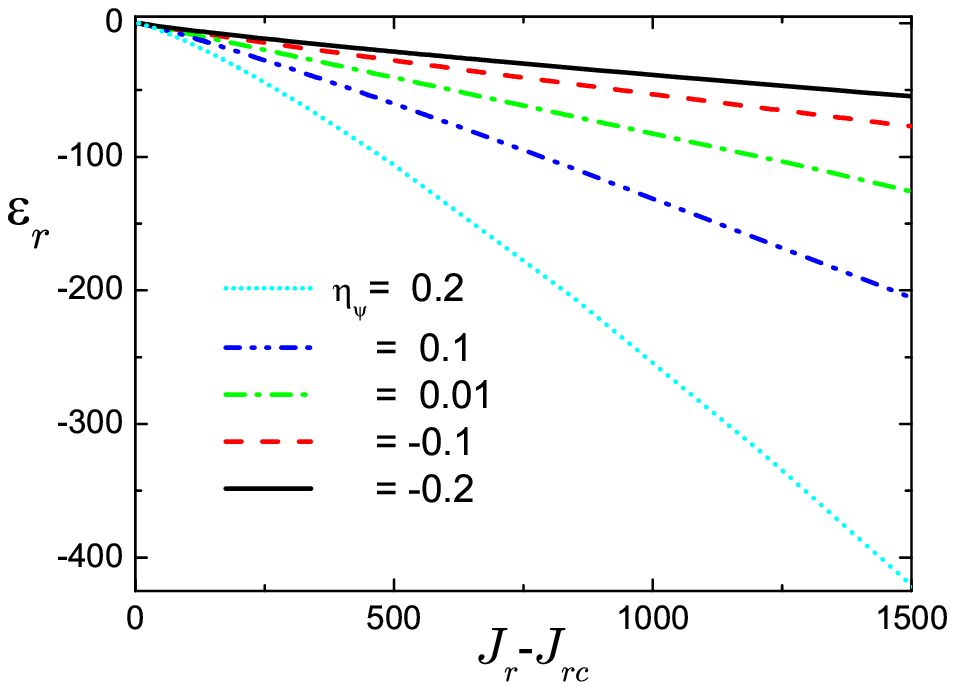} \includegraphics{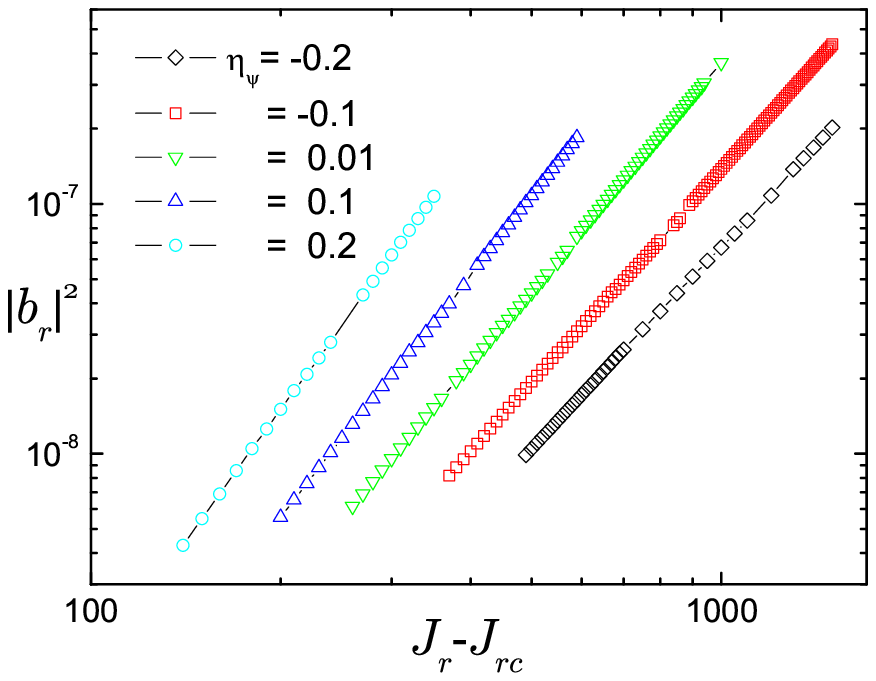} \vspace*{0cm}
\vspace{-0.3cm}
\hspace{-1cm} {\bf\Large (a)}
\hspace{6.7cm}  {\bf\Large (b)}
\caption{
(Color online) Impurity chemical potential, $\epsilon_r$, and hybridization
amplitude, $b_r$, (log-log plot) as a function of the renormalized Kondo
coupling, $J_r$, for various ASL exponents.
Here $\epsilon_r$, $b_r$, and $J_r$ are dimensionless rescaled  variables.}
\label{fig2}
\end{figure}

Solving Eq. (15) numerically, one can find the hybridization
amplitude $|b_r|^{2}$ as a function of the Kondo coupling $J_{r}$.
We show the I-QPT in Fig. 2, where both $b_{r}$ and $\epsilon_{r}$
vanish as $J_{r} \rightarrow J_{rc}$. It is important to notice
that the $x$-axis is $J_{r} - J_{rc}$ instead of $J_{r}$. This
means that the impurity QCP matches the origin of the $x$-axis.
The absolute value of the impurity chemical potential increases
rapidly as the ASL exponent increases from $\eta_{\psi} = - 0.2$
to $\eta_{\psi} = 0.2$ [Fig. \ref{fig2}(a)]. Accordingly, the
increasing ratio of the hybridization order parameter is largest
for $\eta_{\psi} = 0.2$ and smallest for $\eta_{\psi} = - 0.2$.
This may be associated with localization tendency emerging from a
positive exponent. A further analysis finds a scaling behavior of
the hybridization amplitude not only near the impurity QCP, but
also rather away from the QCP, i.e., in the Kondo-screened phase.
Such a scaling behavior even in the Kondo phase seems to arise
from the criticality of the bulk system. From the log-log plot of
Fig. \ref{fig2}(b), we find the scaling relation \bqa &&
|b_{r}|^{2} \sim (J_{r} - J_{rc})^{f(\eta_{\psi})}  \eqa with
$f(\eta_{\psi}) \approx 3+2\eta_{\psi}$, confirming that the slope
of the positive ASL exponent is larger than that of the negative
one.

%%%%%%%%%%%%%%%%%%%%%%%%%%%%%%%%%%%%%%%%%%%%%%%%%%%%%%%%%%%%%%%%%%%%%%%%
%Fig.2
%%%%%%%%%%%%%%%%%%%%%%%%%%%%%%%%%%%%%%%%%%%%%%%%%%%%%%%%%%%%%%%%%%%%%%%%

The I-QPT can be also found in the impurity-spin susceptibility,
\bqa && \chi_{imp} = - \frac{\partial^{2}f_{imp}(h)}{\partial
h^{2}} = - \frac{N_{\sigma}}{\pi} \int_{0}^{\infty}dx
\frac{\epsilon_{r}^{2} - x^{2}\bigl[1 +
\frac{2N_{n}}{\pi\eta_{\psi}}|b_{r}|^{2}\bigl\{(x^{2} +
1)^{\eta_{\psi}/2} -
x^{\eta_{\psi}}\bigr\}\bigr]^{2}}{\bigl(x^{2}\bigl[1 +
\frac{2N_{n}}{\pi\eta_{\psi}}|b_{r}|^{2}\bigl\{(x^{2} +
1)^{\eta_{\psi}/2} - x^{\eta_{\psi}}\bigr\}\bigr]^{2} +
\epsilon_{r}^{2}\bigr)^{2}} . \eqa In the decoupled phase ($J_{r}
< J_{rc}$) the impurity susceptibility diverges in the zero
temperature limit (following the Curie law) while it vanishes in
the screened phase. Since for $\eta_{\psi} = - 0.2$ the hybridization amplitude is smallest,
the impurity-spin susceptibility becomes largest. Approaching the impurity QCP
($J_{r} \rightarrow J_{rc}$), it shows a power-law divergence with
an anomalous critical exponent of the ASL bulk. As shown in
Fig. 3, such curves are well fitted with \bqa && \chi_{imp} \sim
(J_{r} - J_{rc})^{- g(\eta_{\psi})} , \eqa where the scaling
function is $g(\eta_{\psi}) \approx 2 - \eta_{\psi}$. It is
valuable to consider how the behavior of the impurity
susceptibility differs from that of the Pseudogap Kondo
model\cite{PGKM_Susceptibility} which corresponds to the case of
$\eta_{\psi} = 0$.

%%%%%%%%%%%%%%%%%%%%%%%%%%%%%%%%%%%%%%%%%%%%%%%%%%%%%%%%%%%%%%%%%%%%%%%%
%Fig.3
\begin{figure}[t]
\vspace{6cm} \includegraphics{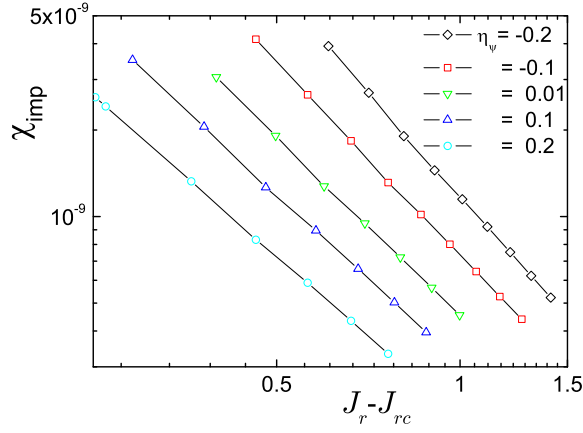} \vspace*{0cm} \caption{ (Color
online) Impurity spin susceptibility (log-log plot) as a function
of the renormalized Kondo coupling for various ASL exponents. }
\label{fig3}
\end{figure}
%%%%%%%%%%%%%%%%%%%%%%%%%%%%%%%%%%%%%%%%%%%%%%%%%%%%%%%%%%%%%%%%%%%%%%%%

%%%%%%%%%%%%%%%%%%%%%%%%%%%%%%%%%%%%%%%%%%%%%%%%%%%%%%%%%%%%%%%%%%%%%%%%
%Fig.4
\begin{figure}[t]
\vspace{6cm} \includegraphics{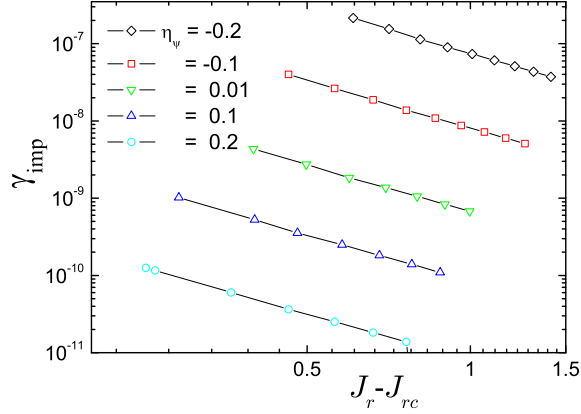} \vspace*{0cm} \caption{ (Color
online) Specific heat coefficient (log-log plot) as a function of
the renormalized Kondo coupling for various ASL exponents. }
\label{fig4}
\end{figure}
%%%%%%%%%%%%%%%%%%%%%%%%%%%%%%%%%%%%%%%%%%%%%%%%%%%%%%%%%%%%%%%%%%%%%%%%

Next, we evaluate the impurity specific heat. The zero temperature
formulation [Eq. (14)] for the impurity free energy can be
transformed to the finite temperature version through the Wick
rotation. Following Refs. \cite{Fradkin_KE_Flux,Doniach}, we find
the impurity free energy at finite temperatures, \bqa && f_{imp} =
N_{\sigma} \int_{-\infty}^{\infty} \frac{d\xi}{\pi}
\frac{1}{e^{\beta_r \xi}+1} \Theta(\xi) +
\frac{2N_{\sigma}N_{n}}{J_{r}} |b_r|^{2} - \epsilon_{r}Q_{\chi}
\nonumber \eqa with a rescaled temperature $\beta_{r}^{-1} = T_{r}
= T/\Lambda$, where the "angle" function $\Theta(\xi)$ is given by
\bqa \Theta(\xi) && =
\tan^{-1}\Bigl(\frac{\frac{2N_{n}}{\pi\eta_{\psi}}|b_r|^{2}
\sin\bigl(\frac{\pi\eta_{\psi}}{2}\bigr)|\xi|^{1+\eta_{\psi}}
}{\xi\bigl[1 + \frac{2N_{n}}{\pi\eta_{\psi}}|b_r|^{2}\bigl\{(-
\xi^{2} + 1)^{\eta_{\psi}/2} -
\cos\bigl(\frac{\pi\eta_{\psi}}{2}\bigr)|\xi|^{\eta_{\psi}}\bigr\}\bigr]
+ \epsilon_{r}} \Bigr)+\frac{\pi}{2}(1-{\rm sign}(\xi)) ~~~
\mbox{for} ~~ |\xi| < 1 , \nn && =
\tan^{-1}\Bigl(\frac{\frac{2N_{n}}{\pi\eta_{\psi}}|b_r|^{2}\sin\bigl(\frac{\pi\eta_{\psi}}{2}\bigr)\xi\bigl\{(\xi^{2}
- 1)^{\eta_{\psi}/2} - |\xi|^{\eta_{\psi}}\bigr\}}{\xi\bigl[1 +
\frac{2N_{n}}{\pi\eta_{\psi}}|b_r|^{2}\cos\bigl(\frac{\pi\eta_{\psi}}{2}\bigr)\bigl\{(\xi^{2}
- 1)^{\eta_{\psi}/2} - |\xi|^{\eta_{\psi}}\bigr\}\bigr] +
\epsilon_{r} } \Bigr)+\frac{\pi}{2}(1-{\rm sign}(\xi)) ~~~~
\mbox{for} ~~ |\xi| \geq 1 . \eqa Here, the denominator and
numerator in the angle function correspond to the real and
imaginary parts of the kernel for the impurity free energy [Eq.
(14)], respectively.

%%%%%%%%%%%%%%%%%%%%%%%%%%%%%%%%%%%%%%%%%%%%%%%%%%%%%%%%%%%%%%%%%%%%%%%%
%Fig.5
\begin{figure}[t]
\vspace{6cm} \includegraphics{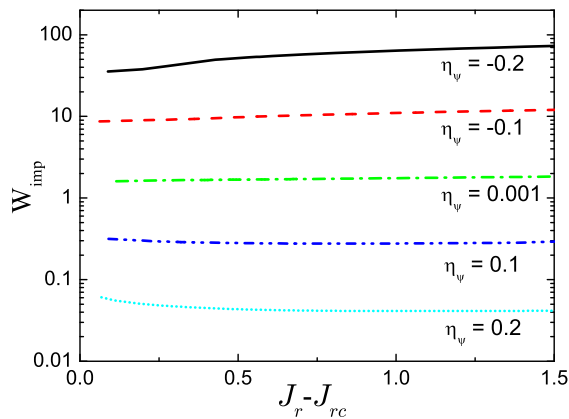} \vspace*{0cm} \caption{ (Color online)
Wilson's ratio (y axis-log plot) as a function of the renormalized
Kondo coupling for various ASL exponents. } \label{fig5}
\end{figure}
%%%%%%%%%%%%%%%%%%%%%%%%%%%%%%%%%%%%%%%%%%%%%%%%%%%%%%%%%%%%%%%%%%%%%%%%

We find the impurity entropy \bqa && S_{imp} = - \frac{\partial
f_{imp}}{\partial T_r}\Bigl|_{\epsilon_{r}, b_r} = N_{\sigma}
\int_{-\infty}^{\infty} \frac{d\xi}{\pi}
\frac{\xi}{T_r}\frac{\partial}{\partial \xi}
\Bigl(\frac{1}{e^{\beta_r \xi}+1}\Bigr) \Theta(\xi) \eqa and
specific heat coefficient \bqa && \gamma_{imp} =
\frac{C_{imp}}{T_r} = \frac{\partial S_{imp}}{\partial T_r} =
\frac{N_{\sigma}}{T_{r}^{2}} \int_{-\infty}^{\infty}
\frac{d\xi}{\pi} \xi^{2} \frac{\partial}{\partial \xi}
\Bigl(\frac{1}{e^{\beta_r \xi}+1}\Bigr)\frac{\partial
\Theta(\xi)}{\partial \xi} . \eqa Taking the zero temperature
limit, we find the self-consistent results in Fig. 4, using the
solutions of Eq. (15). The latter terms in Eq. (21) ensure that
the impurity entropy is $S_{imp}/N_{\sigma} = \ln 2$ in the
decoupled phase, consistent with our expectation. In the Kondo
phase the impurity entropy becomes vanished. But, we note that
more elaborate calculations result in small nonzero entropy
contributions in the Kondo phase.\cite{NCA,Entropy} The
$\gamma_{imp}$ coefficient shows a  behavior similar to the
impurity susceptibility $\chi_{imp}$, diverging as $J_{r}
\rightarrow J_{rc}$. It exhibits the scaling behavior , \bqa &&
\gamma_{imp} \sim (J_{r} - J_{rc})^{-h(\eta_{\psi})} \eqa with
$h(\eta_{\psi}) \approx 2 - 0.2 \eta_{\psi}$ in our numerical
analysis.

Using the impurity susceptibility and specific heat coefficient,
one can find the Wilson's ratio in the zero temperature limit \bqa
&& W_{imp}(T_r\rightarrow 0) =
\frac{\gamma_{imp}}{\chi_{imp}}\Bigl|_{T_r\rightarrow 0} . \eqa In
Fig. 5 we plot this value as a function of the rescaled Kondo
coupling $J_{r} - J_{rc}$.
Remember $W_{imp} = 2$ in the Kondo effect of the Fermi liquid.
Here, we also obtain a similar value for the Pseudogap Kondo model ($\eta_{\psi} = 0.001$).
An important observation is that the Wilson's ratio is strongly
dependent on the ASL exponent. For the negative exponent the
Wilson's ratio becomes enhanced while it gets suppressed for the
positive one. This implies that the Wilson's ratio can be utilized
as a probe for revealing the nature of SLs, more generally,
criticality of the bulk system. This is the main message of the
present study.

\section{Summary and discussion}

It is valuable to remind several assumptions for solving the ASL
Kondo problem. First of all, we have considered effects of
staggered spin fluctuations on dynamics of a magnetic impurity,
ignoring those of uniform spin correlations, since
antiferromagnetic spin fluctuations are most susceptible in the
ASL bulk, thus expected to give dominant contributions on this
problem. In addition, ferromagnetic spin correlations do not show
anomalous scaling, implying that such contributions would coincide
with the Pseudogap Kondo effect, thus not so interesting. Our
second assumption is in writing a spinon Green's function, where
effects of gauge fluctuations are introduced in an anomalous
critical exponent. In the present paper we have used the scaling
exponent as a phenomenological parameter. Both assumptions are
compatible since the critical exponent of the staggered spin-spin
correlation function is consistent with that of the single
particle propagator when both critical exponents are evaluated in
a gauge invariant way.

The third one is rather an approximation than an assumption for
solving the effective impurity action while the above two are
basic assumptions for deriving the impurity action. In the
slave-boson representation of this effective impurity action we
have performed the large-$N_{\sigma}$ analysis introducing the
hybridization order parameter. Although well-defined quasiparticle
excitations do not exist in the case of $\eta_{\psi} \not= 0$, it
was shown that the I-QPT occurs between the local-moment state and
the Kondo-screened phase. Evaluating the impurity spin
susceptibility and specific heat coefficient, we found that the
Wilson's ratio depends strongly on the ASL exponent. This result
has an important physical meaning because the Wilson's ratio for
the magnetic impurity reflects criticality of the bulk system.
This conclusion will be available to general critical systems with
exact screening particulary, where the expression of the Kondo
vertex is the same as that of the present
paper.\cite{Criticality_Kondo} In this respect the Wilson's ratio
for the magnetic impurity may be one possible probe for measuring
bulk criticality.

It is interesting to compare the ASL Kondo physics with the Kondo
effect in the Luttinger liquid,\cite{Furusaki,MunDaeKim} since the
ASL can be considered as the high dimensional realization of the
Luttinger liquid. In the Luttinger liquid the Kondo interaction
term can be decomposed to the forward and backward scattering
channels, analogous to the uniform and staggered ones in the ASL.
It was shown that the forward scattering channel is irrelevant in
the renormalization group analysis up to two-loop
order.\cite{MunDaeKim} Similarly,  in this paper we take into account only the antiferromagnetic
correlation channel  for the ASL Kondo effect,
although it is not proven that the ferromagnetic channel is irrelevant.
The backward impurity scattering in the
Luttinger liquid was shown to cause anomalous scaling, in
particular, power-law behavior of the Kondo temperature owing to
the presence of the anomalous critical exponent in the Luttinger
liquid.\cite{MunDaeKim} This is basically the same as the
ASL-Kondo effect that the ASL criticality results in anomalous
scaling on the impurity physics, although there is no phase
transition in the Luttinger liquid owing to one dimensionality.

It should be noted that the present mean-field analysis is
difficult to describe correct scaling behaviors in the
over-screened phase since the hybridization order parameters are
not regarded as dynamic variables but static ones. This
approximation scheme seems to be more appropriate when
quasiparticle excitations are well defined, thus the conventional
Kondo screening picture is applicable. The slave-boson mean-field
scheme can be improved using the non-crossing
approximation,\cite{NCA} where such hybridization parameters are
taken to be dynamic variables, thus quantum fluctuations are more
involved. Performing the Hubbard-Stratonovich transformation for
the nonlocal (for time) hopping term in Eq. (9), we find an
effective impurity action \bqa && {\cal S}_{NCA} =
\int\frac{dk_{0}}{2\pi} \Bigl[ \sum_{\sigma = 1}^{N_{\sigma}}
\chi_{\sigma}^{\dagger}(ik_{0} - h_{\sigma} +
\epsilon_{\chi})\chi_{\sigma} + \frac{N_{\sigma}}{2J_{K}}
\sum_{n=1}^{N_{n}} b_{n}^{s\dagger}\gamma_{0}b_{n}^{s}  -
\epsilon_{\chi}Q_{\chi} \Bigr] \nn && + \int\frac{dk_{0}}{2\pi}
\Bigl[ \int \frac{dk'_{0}}{2\pi}
\frac{\Sigma_{\chi}(k_0)\Sigma_{b}(k'_0)}{i(k_0-k'_0)F(k_0-k'_0)}
- \sum_{\sigma =
1}^{N_{\sigma}}\chi_{\sigma}^{\dagger}(k_0)\Sigma_{b}(k_0)\chi_{\sigma}(k_0)
+ \sum_{n = 1}^{N_{n}} b_{n}^{s\dagger}(k_0) \Sigma_{\chi}(-k_0)
b_{n}^{s}(k_0)  \Bigr] , \nonumber \eqa where $\Sigma_{\chi}(k_0)$
and $\Sigma_{b}(k_0)$ are fermion and boson self-energies,
respectively, determined by the following self-consistent NCA-type
equations \bqa && \Sigma_{\chi}(\tau'-\tau) = {\cal
F}(\tau-\tau')\langle\sum_{\sigma =
1}^{N_{\sigma}}\chi_{\sigma}^{\dagger}(\tau)
\chi_{\sigma}(\tau')\rangle , \nn && \Sigma_{b}(\tau-\tau') = -
{\cal F}(\tau-\tau')\langle\sum_{n = 1}^{N_{n}}
b_{n}^{s\dagger}(\tau)b_{n}^{s}(\tau') \rangle \nonumber \eqa with
${\cal F}(k_0) = ik_0 F(k_0)$. This kind of approximation is well
known to catch non-Fermi liquid physics in the multi-channel Kondo
model.\cite{NCA} Scaling behaviors of both bosonic and fermionic
self-energies are expected in the low energy limit, causing
anomalous critical physics to this system even in the case of
$\eta_{\psi} = 0$. Inserting expected scaling forms for both the
self-energies and renormalized Green's functions to the NCA
equations, we would obtain the total anomalous scaling exponents
which are expected to be sum of the critical exponents of the
multichannel Pseudogap Kondo model and the ASL scaling dimension
approximately, considering the presence of the ASL scaling
exponent in ${\cal F}(\tau-\tau')$. It will be interesting to
examine how the scaling exponents in the conventional bulk are
affected by the presence of the ASL exponent.

Applying magnetic fields to the ASL, the impurity QPT is expected
to disappear. Because external magnetic fields would result in
finite density of states at the Fermi energy, the conventional
Kondo physics may appear, where only the over-screened Kondo phase
would occur, independent of the Kondo interaction. Furthermore,
gauge fluctuations would be dissipative due to the finite density
of states, and the bulk system becomes more "Fermi liquid"-like,
supporting the above expectation.

In the present analysis we did not consider scattering due to
randomly distributed disorder potentials. One of the present
authors has studied the role of random potentials in the ASL, and
found that such a spin liquid phase remains stable against weak
disorders because massless Dirac spinons at the interacting fixed
point live in higher spatial dimensions than two owing to the
presence of the anomalous critical exponent.\cite{Kim_Disorder}
Remember the presence of the delocalization transition above two
spatial dimensions. However, it is not clear whether the diffusive
nature appears or not in the ASL. If so, the presence of finite density
of states due to random potential scattering may destroy the I-QPT
as the case of magnetic fields.


\begin{thebibliography}{9}
\bibitem{Exponent_Large_N} The renormalized spinon propagator
is expressed as $G^{-1}_{ASL}(k) = G_{0}^{-1}(k) + \Sigma(k)$,
where $G_{0}^{-1}(k) = i\gamma_{\mu}k_{\mu}$ is the inverse of the
bare spinon propagator, and $\Sigma(k)$ the spinon self-energy
resulting from long-range gauge interactions. In the
$1/N_{\sigma}$ expansion the self-energy is represented as
$\Sigma(k) =
\int\frac{d^3q}{(2\pi)^{3}}\gamma_{\mu}G_{0}(k+q)\gamma_{\nu}D_{\mu\nu}(q)$,
where $D_{\mu\nu}(q)$ is the renormalized propagator of the U(1)
gauge field due to polarization of massless Dirac fermions. The
gauge propagator is obtained to be $D_{\mu\nu}(q) \approx
\Pi_{\mu\nu}^{-1}(q) = \frac{8}{N_{\sigma}q}\Bigl(\delta_{\mu\nu}
- \frac{q_{\mu}q_{\nu}}{q^2}\Bigr)$ in the Lorentz gauge, where
$\Pi_{\mu\nu}(q) =
N_{\sigma}\int\frac{d^3k}{(2\pi)^{3}}Tr[G_{0}(k)\gamma_{\mu}G_{0}(k+q)\gamma_{\nu}]$
is the polarization function of Dirac fermions. Inserting this
gauge propagator into the expression of the self-energy, one can
find the spinon self-energy of logarithmic momentum dependence
$\Sigma(k) =
i\eta_{\psi}\gamma_{\mu}k_{\mu}ln\Bigl(\frac{\Lambda}{k}\Bigr)$,
where $\eta_{\psi}$ is the anomalous exponent and $\Lambda$, the
momentum cutoff. The absolute value of the exponent $\eta_{\psi}$
is proportional to the inverse of the flavor number, i.e.,
$|\eta_{\psi}| \sim N_{\sigma}^{-1}$. Such logarithmic momentum
dependence should be considered as the lowest order in the
algebraic function. As a result, one can obtain the algebraically
decaying spinon propagator [Eq. (2)] from the following
nonperturbative consideration $G_{ASL}^{-1}(k) =
i\gamma_{\mu}k_{\mu}\Bigl[1 +
\eta_{\psi}ln\Bigl(\frac{\Lambda}{k}\Bigr)\Bigr] \approx
i\gamma_{\mu}k_{\mu}\Bigl(\frac{\Lambda}{k}\Bigr)^{\eta_{\psi}}$.
\bibitem{PGKM} The saddle-point equations of the Pseudogap Kondo
model can be obtained from Eq. (14) in the limit of $\eta_{\psi}
\rightarrow 0$ \bqa && \frac{1}{J_{r}} = \frac{1}{2\pi^{2}}
\int_{0}^{\infty}dx \frac{x^{2}
\ln\bigl(1+\frac{1}{x^{2}}\bigr)\bigl[1 +
\frac{N_{n}}{\pi}|b_{r}|^{2}\ln\bigl(1+\frac{1}{x^{2}}\bigr)\bigr]}{x^{2}\bigl[1
+
\frac{N_{n}}{\pi}|b_{r}|^{2}\ln\bigl(1+\frac{1}{x^{2}}\bigr)\bigr]^{2}
+ \epsilon_{r}^{2} } , \nn && \frac{Q_{\chi}}{N_{\sigma}} = -
\frac{1}{\pi} \int_{0}^{\infty} dx \frac{
\epsilon_{r}}{x^{2}\bigl[1 +
\frac{N_{n}}{\pi}|b_{r}|^{2}\ln\bigl(1+\frac{1}{x^{2}}\bigr)\bigr]^{2}
+ \epsilon_{r}^{2} } , \nonumber \eqa giving rise to Eq. (16) for
the impurity QCP.
\bibitem{PGKM_Susceptibility} In the Pseudogap Kondo model
the impurity susceptibility is given by \bqa && \chi_{imp} = -
\frac{N_{\sigma}}{\pi} \int_{0}^{\infty}dx \frac{\epsilon_{r}^{2}
- x^{2}\bigl[1 + \frac{N_{n}}{\pi}|b_r|^{2}\ln\bigl(1 +
\frac{1}{x^{2}}\bigr)\bigr]^{2}}{\bigl(x^{2}\bigl[1 +
\frac{N_{n}}{\pi}|b_r|^{2}\ln\bigl(1 +
\frac{1}{x^{2}}\bigr)\bigr]^{2} + \epsilon_{r}^{2}\bigr)^{2}} .
\nonumber \eqa
Remember that the impurity susceptibility based on
the large-$N_{\sigma}$ analysis exhibits a logarithmic correction
due to the upper critical dimensionality.\cite{Fradkin_KE_Flux}
\bibitem{Criticality_Kondo} Note that the O(3) critical bulk
described by spin 1 critical fluctuations is not described by the
present Kondo model. The present results cannot be applied to all
critical bulk systems in spite of their generality because the
vertex in Kondo interaction differs from each system, depending on
quantum numbers of critical modes in the bulk system.
\end{thebibliography}
\end{document}